\documentclass{elsart}
\usepackage[dvips]{graphicx}

\begin{document}

\begin{frontmatter}



\title{Recent progress of avalanche photodiodes in high-resolution\\
 X-rays and $\gamma$-rays detection}


\author[label1]{J. Kataoka},
\corauth[corl]{Corresponding author. Tel.: +81-3-5734-2388; fax: +81-3-5734-2389}
\ead{kataoka@hp.phys.titech.ac.jp}
\author[label1]{T. Saito},
\author[label1]{Y. Kuramoto},
\author[label1]{T. Ikagawa},
\author[label1]{Y. Yatsu},
\author[label1]{J. Kotoku},
\author[label1]{M. Arimoto},
\author[label1]{N. Kawai},
\author[label2]{Y. Ishikawa} and
\author[label2]{N. Kawabata}
\address[label1]{Tokyo Institute of Technology, 2-12-1 Ohokayama, 
Meguro, Tokyo, 152-8551, Japan}
%$\address[label2]{Hiroshima University, Hiroshima, Japan}
\address[label2]{Hamamatsu Photonics K.K., Hamamatsu, Shizuoka, Japan}

\begin{abstract}
We have studied the performance of large area avalanche photodiodes 
(APDs) recently developed by Hamamatsu Photonics K.K, in 
high-resolution X-rays and $\gamma$-rays detections. 
We show that reach-through APD can be an excellent soft 
X-ray detector operating at room temperature or moderately cooled 
environment. We obtain the best energy resolution 
ever achieved with APDs, 6.4 $\%$ for 5.9 keV X-rays, and obtain the 
energy threshold as low as 0.5 keV measured at $-$20 $^{\circ}$C.
Thanks to its fast timing response, signal carriers in the APD device
 are collected within a short time interval of 1.9 nsec (FWHM).
This type of APDs can therefore be used as a low-energy, 
high-counting particle monitor onboard the forthcoming Pico-satellite 
Cute1.7. As a scintillation photon detector, reverse-type APDs have a good 
advantage of reducing the dark noise significantly. The best FWHM 
energy resolutions of 9.4$\pm$0.3$\%$ and 4.9$\pm$0.2$\%$ were obtained  
for 59.5 keV and 662 keV $\gamma$-rays, respectively, 
as measured with a CsI(Tl) crystal. Combination of APDs with various 
other scintillators (BGO, GSO, and YAP) also showed better results 
than that obtained with a photomultiplier tube (PMT). 
These results suggest that APD could be a promising device for replacing 
traditional PMT usage in some applications. In particular 2-dim 
APD array, which we present in this paper, will be a promising device
for a wide-band X-ray and $\gamma$-ray imaging detector in future 
space research and nuclear medicine.
\end{abstract}

\end{frontmatter}
\section{introduction}
In recent years avalanche photodiodes (APD) have attracted considerable 
attention since good features of both photodiodes (PDs) and 
photomultiplier tubes (PMTs) are shared by 
APDs\cite{web74}. In fact, APDs have the quantum efficiency 
close to 100$\%$ in the visible and near infrared, can be very 
compact and less affected by magnetic field, and  produces an 
internal gain of 10$-$100 or more, though it is much less than typical 
PMT gain. Thus the basic properties of APD is well suited to read 
out small numbers of photons, so long as it has large detection area 
and is operated under stable conditions.  

For a long time, however, APDs were limited to very small surfaces, 
and mainly used as a digital device for light communications 
(e.g., a receiver for optical fibers). During the past decade, 
a large area APDs operating as a linear detector has also been
available. As a scintillation photon detector, 
Moszy\'{n}ski et al.\cite{mos98,mos01,mos03} have obtained a better or 
comparable energy 
resolution to those observed with a PMT. Moreover, operations of 
APDs at low temperature reduce the dark current noise contribution. 
This significantly improves the sensitivity to low-intensity signals, 
such as weak scintillation light produced by low energy X-rays. 

In this paper, we report the performance of large area APDs recently 
developed by Hamamatsu Photonics K.K to determine its suitability 
as a low energy X-rays and $\gamma$-rays scintillation detector.
After recalling APD structures, we summarize fundamental 
properties  of three different APDs in $\S$2. In $\S$3, 
we present the performance of reach-through APD in direct detection of 
soft X-ray photons. In $\S$4, we show the energy spectra of $\gamma$-ray
sources measured with four different scintillators coupled to the 
reverse-type APDs. As a future imaging application, 
the performance of a pixel APD array will be presented in $\S$5. 
Finally we summarize our results in $\S$6. 

\section{APD Structures and parameters}
\subsection{APD types}
Three different types of APDs are now commercially available:  
(a) ``beveled-edge'', (b) ``reach-through'', and (c) ``reverse-type'' 
diode (Figure 1). Structure (a), the  ``beveled-edge'' diode is a traditional
p$^{+}$n junction in which the n-type resistivity is chosen so as to 
make the breakdown voltage very high (typically 2000 V)
\cite{mos98,mos01,mos03,och96,mos99,mos02}.

\begin{figure}[htb]
\begin{center}
\vspace*{0mm}
\includegraphics[width=7.8cm,angle=0,clip]{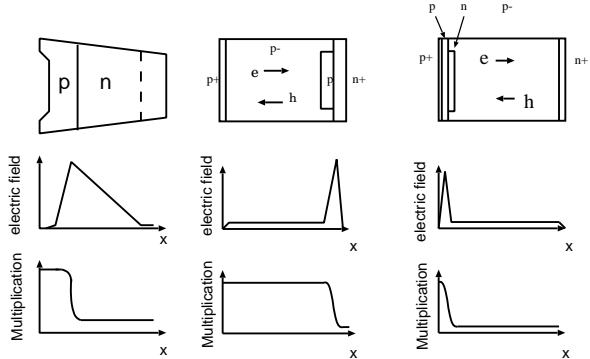}
\caption{Internal structures of three different types of APDs.
(a)beveled-edge, (b)reach-through, and (c)reverse type (from left to
 the right). The electric field profiles and  the gain profiles in 
each APDs are also shown in the middle and bottom panels.}
\end{center}
\end{figure}

Structure (b), ``reach through'' type, applies to a diode in which the 
depletion layer comprises both a relatively wide drift region of fairly 
low field ($\sim$ 2 V/$\mu$m) and a relatively narrow region of field 
sufficient for impact ionization (25$-$30 V/$\mu$m). The advantage 
of such a structure is that only relatively low voltages 
are required to full depleting the devices. 
Traditional reach-through APDs have a wide low-field drift region
($\sim$ 100 $\mu$m) at the front of the device, with the multiplying 
region at the back. A disadvantage is that most of the 
dark current undergoes electron multiplication, resulting that 
large area devices tend to be somewhat noisy.

\begin{table}[h]
\caption{Parameters for Hamamatsu APDs} 
\label{tab:fonts}
\begin{center}       
\begin{tabular}{l|r r r} %% this creates two columns
\hline
\rule[-1ex]{0pt}{3.5ex}  Name &\hspace{2mm} SPL 2407 &\hspace{2mm} S8664-55 &\hspace{2mm}S8664-1010\\
\hline
\rule[-1ex]{0pt}{3.5ex}  Type & (b) & (c)  & (c)\\
\rule[-1ex]{0pt}{3.5ex}  Surface area\hspace{2mm} &\hspace{2mm}$\phi$ 3mm&\hspace{2mm} 5$\times$5mm$^2$
 &\hspace{2mm} 10$\times$10mm$^2$ \\
\rule[-1ex]{0pt}{3.5ex}dark current:  $I_{\rm D}$  & 4.4 nA &
 0.4 nA &  1.7 nA\\
\rule[-1ex]{0pt}{3.5ex}capacitance:  $C_{\rm det}$  & 10.2 pF &
 88 pF &  269 pF \\
\rule[-1ex]{0pt}{3.5ex}breakdown:  V$_{\rm brk}$ & 647 V &
 390 V &  433 V \\  
\hline
\end{tabular}
\end{center}
\end{table} 

 The ``reverse type (c)'' is specifically designed to couple with 
scintillators. This type is quite similar to the reach-through APD, 
but the narrow high-field multiplying region has been moved to the 
front end, typically about 5 $\mu$m from the surface of the 
device\cite{lec99,mci96}. Since major scintillators emit at 
wavelength of 500 nm or less, most of lights from scintillators are  
absorbed within the first 1$-$3 $\mu$m of the depletion layer and 
generates electrons which undergo full multiplication. Whereas most 
of the dark current undergoes only hole multiplication, reducing the 
noise contribution significantly. 

In this paper, we summarize the basic properties of APDs most recently 
developed by Hamamatsu (see Table 1): SPL2407, S8664-55, and S8664-1010. 
SPL2407 is a reach-through type, whereas S8664-55 and S8664-1010 
are the reverse-type APDs. SPL 2407 ($\phi$ 3mm) has a depletion layer of 
130 $\mu$m thickness, and can be used in direct detection of soft X-rays 
below 20 keV. 

\subsection{Gain and dark current}
The gain characteristic of APDs can be measured under constant
illumination of monochromatic light source recording the photocurrent 
of the APD as a function of bias voltage. We use a light emitting diode 
(LED) producing light signals of 525 nm. At voltages lower than 50 V
(10 V for SPL 2407), the APD gain can be regarded as unity since the 
photocurrent remained constant. Figure 2 shows variations of 
APD gain as a function of bias voltage, measured at +20$^{\circ}$C.  
At a gain of 30, the gain variations on bias voltage are approximated by 
+2.7 $\%$/V for reverse-type APDs (S8664-55, S8664-1010) 
whereas +0.5 $\%$/V for reach-through APD (SPL2407). 
Note that this is comparable to the voltage coefficient of typical PMTs 
($\sim$ +2 $\%$/V).

\begin{figure}[htb]
\begin{center}
\vspace*{0mm}
\includegraphics[width=6.0cm,angle=90,clip]{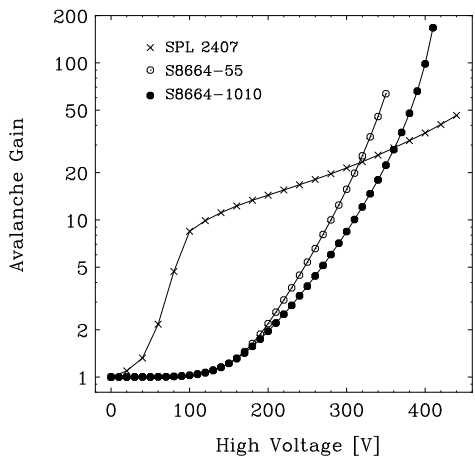}
\caption{Gain variations of Hamamatsu APDs 
measured at room temperature (+20$^{\circ}$C).}
\end{center}
\end{figure}

As discussed in detail in Ikagawa et al.\cite{ika03} 
APD gain also depends on temperature. At a gain of 50, 
gain variations of APD ranges in a few 
$\%$/$^{\circ}$C, which is an order of magnitude larger than typical PMTs.
Therefore, temperature control could be more critical problems for APDs.
Throughout this paper, temperature was controlled in a thermostat within 
0.1$^{\circ}$C. The corresponding deviation in gain is less than 0.3 $\%$.

The leakage currents were measured at room temperature:  
4.4 nA, 0.4 nA, and 1.7 nA at a gain of 30  for SPL 2407, S8664-55 and 
S8664-1010, respectively. There values are extremely low compared 
to those reported for the beveled-edge APDs of a similar size. For example 
Moszy\'{n}ski et al.\cite{mos98}  reported that 
leakage current of $\phi$ 16 mm APD  is $\ge$100 nA at room temperature. 
Regarding the Hamamatsu APDs, 
leakage current further decreases to 10$-$100 pA  level at 
$-$20 $^{\circ}$C.  

\section{Performance as a soft X-ray detector}
\subsection{Energy spectra}
A reach-through APD of a 130 $\mu$m thickness can potentially 
detect soft X-rays below 20 keV with efficiencies 
greater than 10$\%$.  
The signal amplification in the APD devices has a good advantage of 
detecting soft X-rays even at room temperature or at lightly 
cooled environment\cite{yat04,kat04}. Figure 3  presents the 
energy spectrum of 5.9 keV X-rays from a $^{55}$Fe source 
measured with SPL 2407 at $-$20$^{\circ}$C. Note that energy threshold 
is as low as $E_{\rm th}$ $\sim$ 0.5 keV. The K-shell peaks of Mn
K$_{\alpha}$ and K$_{\beta}$ are marginally resolved in the line
profile. The FWHM width of the 5.9 keV peak was $\Delta$E $\sim$ 379 eV 
(6.4 $\%$; Figure 3), which is the best record ever achieved with 
APDs. This resolution is clearly better than those
obtained with the proportional counters.

\begin{figure}[htb]
\begin{center}
\vspace*{0mm}
\includegraphics[width=6.0cm,angle=90,clip]{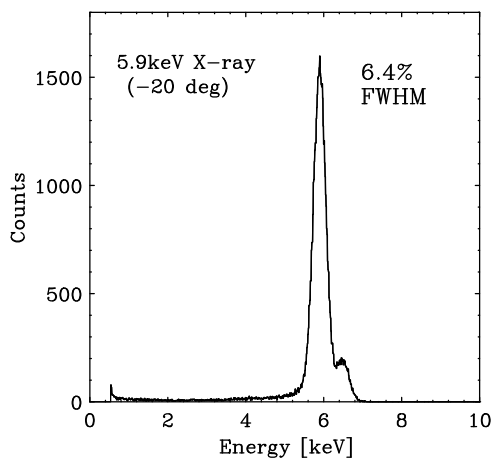}
\caption{Energy spectrum of 5.9 keV X-rays at $-$20$^{\circ}$C.}
\end{center}
\end{figure}

\subsection{High-rate counting}
Fast timing is another excellent property of APDs.
It has been reported that an APD with a few mm$^2$ detection area has 
fast timing properties better or comparable to that of a fast PMT
\cite{kis98,kis01}. We found that signal carriers in the APD device,
SPL 2407, are collected within a short time interval of 1.9 nsec (FWHM). 
Figure 4 presents the output count rate as a function of the input 
(observed) photon rate for 8.0 keV photons. When setting a threshold 
as low as 1 keV for these 1.9 nsec (FWHM) pulses, more than 34 $\%$  
signals were successfully recorded at the maximum input rate of 
$\sim$ 3$\times$10$^8$ cts/s. Note that the reduction of observed 
count rate is exactly consistent with that expected from the 
Poisson distribution with a dead time $\tau$ $\simeq$ 3.6 nsec 
(about twice of 1.9 nsec), and is not due to the  saturation of 
readout electronics. Since the noise level of the APD detector is 
less than 10$^{-2}$ cts/s, we obtain a dynamic range of more than 10$^9$. 

\begin{figure}[htb]
\begin{center}
\vspace*{0mm}
\includegraphics[width=6.0cm,angle=90,clip]{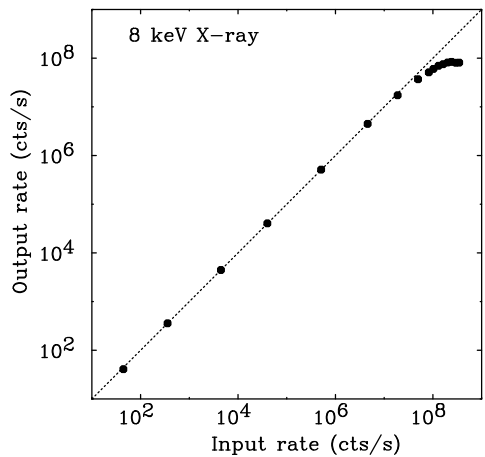}
\caption{Output count rate as a function of the input (observed) photon rate
for 8.0 keV X-rays.}
\end{center}
\end{figure}

Since this type of APD (reach-through type) can also work as a charged 
particle detector, we plan to use it as a high-counting particle  
monitor onboard the forthcoming Pico-satellite Cute1.7\cite{kur04}.
This will be the first mission of using APDs in space as a 
scientific instrument, and study the distribution of low-energy 
($E$ $\le$ 30 keV) electrons and protons trapped 
in the South Atlantic Anomaly (SAA) and aurora band.

\section{Performance as a scintillation photon detector}
\subsection{Read-out of various scintillators}  
We study the performance of reverse-type APD, S8664-1010,  
as a $\gamma$-ray detector coupled with four different 
scintillators; CsI(Tl), BGO, GSO(Ce) and  YAP(Ce). 
A size of the crystals was 10$\times$10$\times$10 mm$^3$, 
and can fully match the sensitive area of the APD. 
Figure 5 compares the pulse height  spectra for 662 keV 
$\gamma$-rays from a $^{137}$Cs source, measured
with a CsI(Tl) crystal at room temperature (+20$^{\circ}$C). Thanks to
its  high quantum efficiency of more than 80$\%$, an excellent FWHM 
energy resolution of 4.9$\pm$0.2 $\%$ was obtained for the APD
($upper$). This is much better than that obtained with the 
1-inch PMT ($lower$:  5.9$\pm$0.1 $\%$ FWHM; Hamamatsu R7899EG), where   
the quantum efficiency of R7899EG is more than 25 $\%$ between 350 nm 
and 450 nm and the signal gain is $\sim$ 1$\times$10$^6$ at an
operating bias voltage of 1000 V. 
More strikingly, APDs' internal gain reduces electric noise contribution 
significantly, resolving K-shell X-ray peak at 32 keV with the energy 
resolution of 20.6$\pm$0.2$\%$ (23.0$\pm$0.1 $\%$ for the PMT)\cite{ika04}. 

\begin{table}[h]
\caption{FWHM energy resolutions for 662 keV $\gamma$-rays} 
\label{tab:fonts}
\begin{center}       
\begin{tabular}{l|c c| c } %% this creates two columns
\hline
\rule[-1ex]{0pt}{3.5ex}  Crystal &\hspace{2mm}APD(+20$^{\circ}$C)  & \hspace{2mm}APD($-$20$^{\circ}$C)  & \hspace{2mm}PMT(+20$^{\circ}$C) \\
\hline
\rule[-1ex]{0pt}{3.5ex}  CsI(Tl) &  4.9$\pm$0.2$\%$  & 5.9$\pm$0.1$\%$  &
 5.9$\pm$0.1$\%$   \\
\rule[-1ex]{0pt}{3.5ex}  BGO     &  8.3$\pm$0.2$\%$  & 7.1$\pm$0.2$\%$   &
 10.4$\pm$0.1$\%$ \\
\rule[-1ex]{0pt}{3.5ex}  GSO(Ce) &  7.8$\pm$0.2$\%$  & 7.1$\pm$0.2$\%$  &  
9.3$\pm$0.1$\%$  \\
\rule[-1ex]{0pt}{3.5ex}  YAP(Ce) &  11.3$\pm$0.3$\%$  & 10.7$\pm$0.2$\%$  &  
12.4$\pm$0.1$\%$ \\
\hline
\end{tabular}
\end{center}
\end{table} 

Similarly, the large area APD is superior to the PMT when coupled with 
various scintillators (BGO, GSO(Ce), YAP(Ce)) as listed in Table 2. 
A good energy resolution of 7.1$\pm$0.2 $\%$ was obtained for 
662 keV $\gamma$-rays, as measured with BGO crystal at
$-$20$^{\circ}$C. The minimum detectable energy was 11.3 keV. 

\begin{figure}[htb]
\begin{center}
\vspace*{0mm}
\includegraphics[width=6.0cm,angle=90,clip]{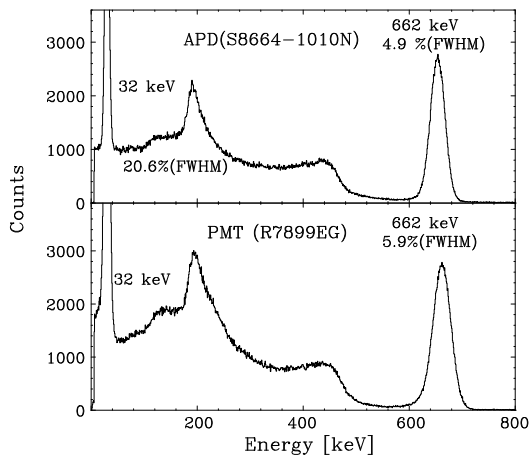}
\caption{Energy spectrum of $^{137}$Cs obtained with CsI(Tl) crystal  
coupled to a large area APD (S8664-1010: $top$) and PMT (R7899EG:
 $bottom$), measured at +20$^{\circ}$C. }
\end{center}
\end{figure}

\subsection{Low energy scintillation detection}
Low leakage current of the reverse-type APD  should have an
excellent advantage for the detection of a low level of scintillation 
light, corresponding to $\gamma$-ray energy below 100 keV. 
To demonstrate the advantage of reverse-type APDs, 
we measured a CsI(Tl) crystal (5$\times$5$\times$5 mm$^3$) 
which can fully match the sensitive area of the S8664-55.

\begin{figure}[htb]
\begin{center}
\vspace*{0mm}
\includegraphics[width=6.0cm,angle=90,clip]{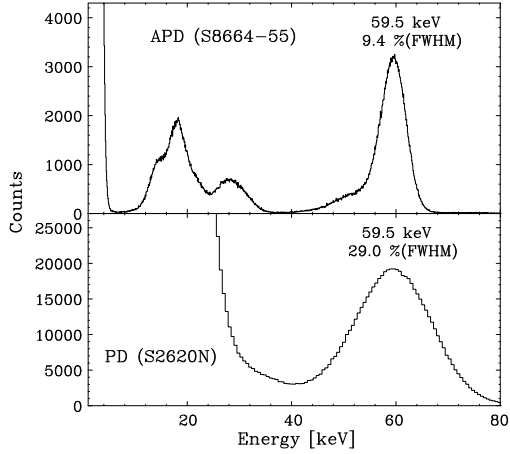}
\caption{Energy spectra of 59.5 keV $\gamma$-rays from a $^{241}$Am
 source measured with a CsI(Tl) crystal coupled to the APD (S8664-55: 
$upper$) and the PD (S2620N).}
\end{center}
\end{figure}

Figure 6 shows the pulse height spectrum of 59.5 keV $\gamma$-rays 
from an $^{241}$Am source, measured at room temperature\cite{ika03}.
The pulse height spectra, using the same CsI(Tl) 
scintillator coupled to the PIN-photodiode (Hamamatsu S2620N-1771: 
5$\times$5 mm$^2$ surface) is also shown for comparison. 
A combination of 14$-$21 keV lines of Np ($L_{\alpha}$, 
$L_{\beta}$ and $L_{\gamma}$) is clearly resolved for APD whereas noise 
dominates for PIN-PD.  Energy resolutions of 59.5 keV $\gamma$-rays 
are 9.4$\pm$0.3 $\%$ for the APD and 29.0$\pm$0.2 $\%$ for the PIN-PD, 
respectively.  These results are one of the best records ever achieved 
with scintillation detectors. The minimum detectable energy is as low 
as 4.6 keV at room temperature, and improves significantly to 1.1 keV 
when cooled at $-$20$^{\circ}$C.

\section{Applications: 32ch APD array}  

Finally, we tested a pixel APD array which offer new design 
options for physics experiments and nuclear medicine, such as imaging devices 
for positron emission tomography (PET). Hamamatsu S8550 (reverse-type) 
is a monolithic 8$\times$4 pixels structure with a surface area of 
2$\times$2 mm$^2$ for each pixel. The common cathode and the individual 
anode of the 32 diodes are connected at the backside of the carrier plates. 
We are testing the performance of S8550, and are developing 
read-out electronics for imaging purposes. Initial results are 
also found in literature\cite{kat04,kap03}.

\begin{figure}[htb]
\begin{center}
\vspace*{0mm}
\includegraphics[width=6.0cm,angle=90,clip]{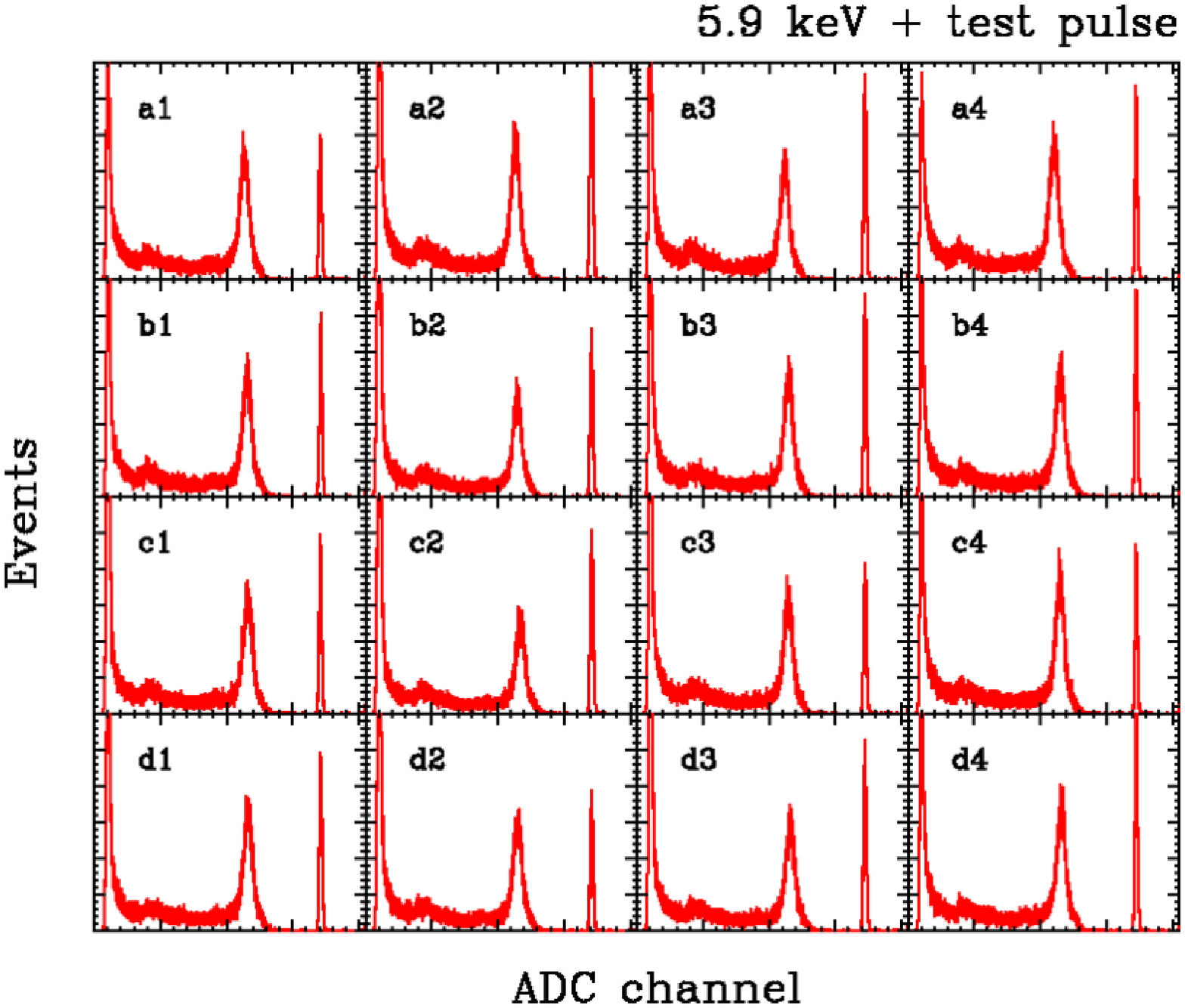}
\includegraphics[width=6.0cm,angle=90,clip]{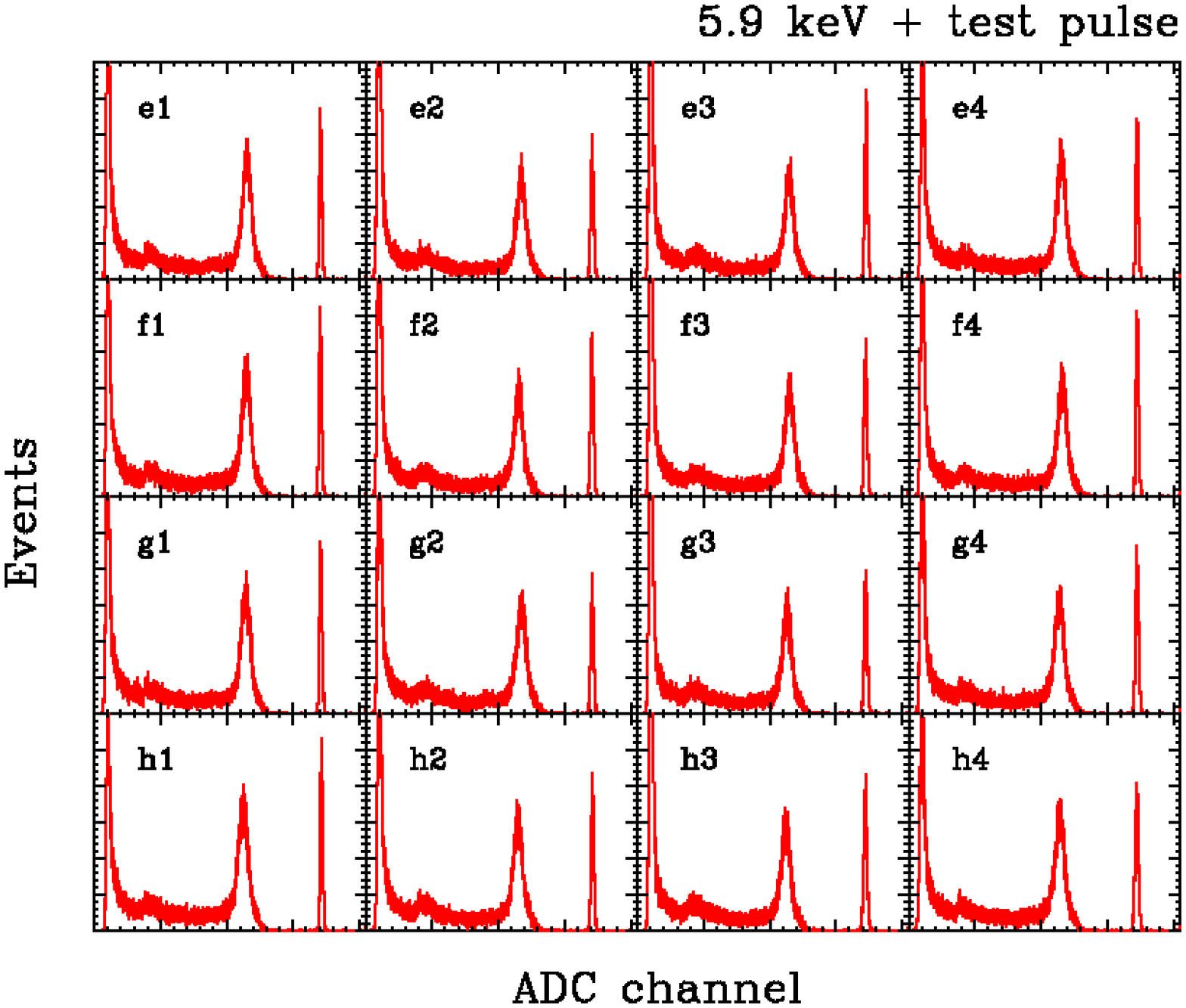}
\caption{A matrix of $^{55}$Fe (5.9 keV) spectra taken with 
each individual pixels of APD array.}
\end{center}
\end{figure}

The leakage current of S8550 is quite uniform between 16 pixels: 
1.4$-$1.9 nA at a gain of 50, measured at $\lambda$ = 420 nm. Capacitance of 
each pixel ranges in 9$-$11 pF. Figure 7 presents energy spectra obtained for 
32 pixels of S8550 using 5.9 keV X-rays. The pixel-to-pixel gain 
non-uniformity was measured to be less than $\pm$3$\%$ at a gain of 50.  
Also one can clearly see that the energy threshold is 
as low as 0.8 keV for each pixels even at a room temperature. Kapusta et al.\cite{kap03}  reported 
that highest crosstalk between adjacent pixels was 4$\%$ at 
a device gain of 60.  These reports 
promise the applicability of Hamamatsu APD array in nuclear medicine 
and space experiments near future. We are developing a 
$\gamma$-ray imaging detector based on the pixel APD arrays coupled to 
CsI(Tl) crystals, and results will be summarized in Saito et al\cite{Sai04}.

\section{Conclusion}
We have studied the performance of large area APDs recently developed by 
Hamamatsu Photonics K.K. 
We showed that reach-through APD can be an excellent soft X-ray detector 
operating at room temperature or moderately cooled environment. 
We obtain the best energy resolution ever achieved with APDs, 
6.4 $\%$ for 5.9 keV X-rays, and obtain the energy threshold as low as 
0.5 keV measured at $-$20 $^{\circ}$C. As a scintillation photon 
detector, reverse-type APDs have an great advantage of reducing the dark 
noise  significantly. We obtain the best FWHM energy resolutions of 
4.9$\pm$0.2$\%$ and 9.4$\pm$0.3$\%$ for 662 keV and 59.5 keV
$\gamma$-rays, respectively, as measured with a CsI(Tl) crystal. 
Combination of APDs with various other scintillators (BGO, GSO, and YAP) 
also showed better results than those obtained with the PMT. 
These results suggest that APD can be a promising device for 
replacing traditional PMT usage in some applications. In particular, 
pixel APD arrays offer new design options for future imaging devices.

\label{}

% \appendix

% \section{}
% \label{}

\end{document}